\documentclass[prl,aps,twocolumn,showpacs]{revtex4}
\usepackage{graphicx}
\usepackage{times}
\usepackage{dcolumn}
\usepackage{natbib}
\begin{document}
\title{All-electron theory of the coupling
between laser-induced coherent phonons in bismuth}
\author{Eeuwe S.\ Zijlstra}
\email[]{zijlstra@physik.uni-kassel.de}
\author{Larisa L.\ Tatarinova}
\author{Martin E.\ Garcia}
\affiliation{Theoretische Physik, Universit\"at Kassel, 
Heinrich-Plett-Str.\ 40, 34132 Kassel, Germany}
\date{\today}
\begin{abstract}
Using \textit{first principles}, all-electron calculations and dynamical 
simulations we study the behavior of the $A_{1g}$ and $E_g$ coherent phonons 
induced in Bi by intense laser pulses. 
We determine the potential landscapes in the laser heated material and show 
that they exhibit phonon-softening, phonon-phonon coupling, and 
anharmonicities. 
As a consequence the $E_g$ mode modulates the $A_{1g}$ oscillations and 
higher harmonics of both modes appear, which explains recent isotropic 
reflectivity measurements. 
Our results offer a unified description of the different 
experimental observations performed so far on bismuth. 
\end{abstract}
%
%
%
\maketitle

Intense femtosecond laser pulses can induce a nonequilibrium state which leads
to sudden and dramatic changes in the potential energy surfaces of different 
solids \cite{Stampfli94}.
This property can be used to excite and manipulate coherent lattice vibrations,
as has been recently shown by several experiments
\cite{Zeiger92,Hase03,Garrett96,Bargheer04} and simulations \cite{Jeschke04}.
In the last years, a variety of experiments on laser excitation of coherent 
phonons has been performed on bismuth 
\cite{Sokolowski03,Hase02,Misochko03,Murray05}, which is a particularly 
interesting solid since its ground-state structure exhibits a Peierls 
distortion.
From the different studies done so far, a number of fundamental aspects still 
remain unexplained, like the detection of higher harmonics \cite{Misochko03}
and the appearance of modes that  are forbidden by symmetry in isotropic 
reflectivity measurements \cite{Hase02}. 

In this letter we perform \textit{ab initio} calculations which show, for the 
first time, the existence of a coupling between laser-induced phonon modes of 
different symmetry.  In addition, we demonstrate that the large amplitude 
atomic vibrations excited by  femtosecond laser pulses are affected by the 
anharmonic part of the potential energy surface, which creates overtones.  

\begin{figure}
  \includegraphics[angle=270,width=6cm]{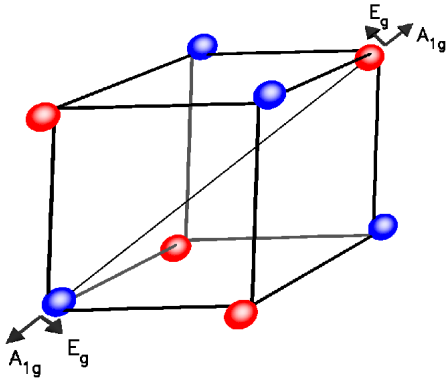}
  \caption{\label{fig_xyz}(Color online) 
  Structure of Bi (see text).
  In the $A_{1g}$ phonon mode the atoms move in the direction of the thin line
  as indicated for two atoms (long arrows).
  In the $E_g$ phonon modes the atoms move in the perpendicular plane
  (e.g., in the directions of the short arrows).}
\end{figure}

The structure of Bi can be derived from a simple cubic atomic packing in two 
steps.
First a simple cubic lattice is deformed by elongating it along one of the 
body diagonals, which  is indicated by a thin line in Fig.\ \ref{fig_xyz}.
A Peierls instability then causes the atoms to be displaced along the 
same diagonal, in opposite directions (Fig.\ \ref{fig_xyz}).
The magnitude of the displacements is determined by the detailed 
interaction between the Bi nuclei and the electrons.
When the electrons are heated
with a laser, the equilibrium positions of the atoms change.
If the duration of the laser pulse is short in comparison with the timescale of
the atomic motion, the ions are lifted at their original equilibrium positions
to the potential energy surface that is created by the hot electrons, from
where they start to swing about their new equilibrium positions 
\cite{Zeiger92}.
This motion of the atoms corresponds to the coherent excitation of the
displacive $A_{1g}$ phonons (Fig.\ \ref{fig_xyz}).
Coherent $E_g$ phonons, which correspond to the motion of the atoms in the 
plane perpendicular to the elongated body diagonal, are also excited by a 
laser, through Raman scattering \cite{Garrett96}.
In Bi the existence of laser-excited coherent $E_g$ phonons has been 
demonstrated through electro-optical measurements \cite{Hase96}.

Recent experiments have shown three exciting results:
(\textit{i}) Through a measurement of the geometrical structure factors of two
x-ray diffraction peaks the most accurate value so far of the maximal 
displacement of the atoms due to the interaction with a short 
laser pulse has been determined \cite{Sokolowski03}.
(\textit{ii}) In isotropic reflectivity measurements apart from the softened 
$A_{1g}$ phonon mode ($2.52$ THz) a signal at lower frequency ($1.61$ THz) 
that can be attributed to the $E_g$ phonons has been detected \cite{Hase02}.
In addition there is also a peak at $3.44$ THz \cite{Hase_private}.
(\textit{iii}) In another study  higher harmonics of the softened $A_{1g}$ 
phonon frequency have been detected \cite{Misochko03}. 

It is the aim of this letter to provide a theoretical explanation of these
experimental results.
To achieve this we performed a dynamical simulation for the $E_g$ and $A_{1g}$ 
degrees of freedom.
The potential energy surface on which the atoms moved was calculated with the 
all-electron full-potential linearized augmented plane wave (LAPW) computer 
program \textsc{WIEN}2k \cite{Blaha01}, which is based on density functional 
theory (DFT).

For our ground state DFT calculations we used the experimental unit cell
parameters of Bi (spacegroup R3m, $a = 4.5332$ {\AA}, $c = 11.7967$ {\AA}).
The Bi atoms occupied the 6c sites (0,0,$z$), where $z$ gives the position of 
the atoms along the above-mentioned diagonal.
We performed DFT calculations for $z = 0.230, 0.231, \dots, 0.250$ $c$, where
$z = 0.250$ $c$ represents the structure in the absence of the atomic 
displacement due to the Peierls instability and $z = 0.234$ $c$ is the 
experimental equilibrium structure.
Based on the total energies for these $z$ we could perform a simulation of the 
coherent $A_{1g}$ phonons.
The degrees of freedom corresponding to the $E_g$ phonons ($x$ and $y$) were 
taken into account by varying $x = 0.000, 0.001, \dots, 0.010$ $a$ and 
$z = 0.234, 0.235, \dots, 0.250$ $c$.
The potential energy surface, on which the ions moved, was then fitted to
the function
\begin{equation}
E_\mathrm{tot} = \sum_{p=0}^4 a_p (z - 0.25)^{2 p} + \left( x^2 + y^2 \right)
\sum_{p=0}^2 b_p (z - 0.25)^{2 p},
\label{eq_surface}
\end{equation}
where the last term describes a coupling between the motion in the $z$ and the
$x$ ($y$) directions.
The fact that it is proportional to $(x^2 + y^2)$ can be easily understood from
the symmetry of the crystal structure, since the coupling cannot depend on the
sign of the displacement (see Fig.\ \ref{fig_xyz}).
Other important details of our DFT calculations were as follows.
LAPW's with energies up to $12.0$ Ry were included in the basis.
Inside the atomic spheres (with radii of $1.376$ {\AA}) additional 5\textit{d},
6\textit{s}, 6\textit{p}, and 6\textit{d} local orbitals were used.
Spin-orbit coupling was treated in a second variational procedure, where the 
scalar relativistic eigenstates up to $3.0$ Ry and local 
6\textit{p}$_{1/2}$ orbitals were used as a basis for the relativistic
calculation.
The entire Brillouin zone was sampled with $512$ $\mathbf{k}$ points using
temperature smearing ($T_e = 1$ mRy).

At elevated temperatures we assumed that there was no exchange of heat between 
the electrons and the ions. 
As a consequence the entropy of the electrons was a constant of motion 
\cite{Stampfli90}.
We further assumed that the electrons were perfectly thermalized at all times 
due to electron-electron interactions.
Therefore the occupation numbers of the Kohn-Sham states were always given by 
a Fermi-Dirac distribution.
Recently, another approach has been proposed 
\cite{Tangney99,Tangney02,Murray05}, where the occupation numbers for the 
electrons and holes of laser-excited Te \cite{Tangney99,Tangney02} and Bi 
\cite{Murray05} are modeled with two temperature distributions using different 
chemical potentials for the electrons and the holes while keeping the number 
of electron-hole pairs constant. 
Although this so-called two-chemical-potential model might work well for 
semiconductors we do not believe that it is appropriate for Bi at high 
laser fluences as we discuss below.

To compare the predictions of the two-chemical-potential model and the model
where the constant entropy of the electrons was the only constraint
we performed additional calculations assuming two chemical potentials.
In agreement with \cite{Murray05} we chose the electron and hole temperatures
$T_e = T_h = 0.5$ eV at $z = 0.234$ $c$.
No heat was assumed to be exchanged between the electrons or the holes and 
the ions.
Therefore the entropy of the electrons and holes were constants of motion.
In this respect we did not follow the method of 
\cite{Tangney99,Tangney02,Murray05}, where the temperatures of the electrons 
and the holes have been kept constant, while the atoms moved on the total 
energy surface, because it yields incorrect forces 
\cite{Wentzcovitch92}.

The potential energy surface $E_\mathrm{tot}(T_e)$ on which the ions moved at 
elevated electronic temperatures was calculated from
\begin{equation}
E_\mathrm{tot}(T_e) = E_\mathrm{tot}( \mathrm{gs}) + \Delta E_\mathrm{band},
\label{eq_nscf}
\end{equation}
where $E_\mathrm{tot}(\mathrm{gs})$ was the self-consistent total energy of
the electronic ground state and $\Delta E_\mathrm{band} = E_\mathrm{band}(T_e) 
- E_\mathrm{band}(\mathrm{gs})$.
This approach is based on the interpretation of the Kohn-Sham energies as 
single-electron excitation energies.
In standard temperature-dependent DFT the electronic occupation numbers are 
incorporated in the self-consistent cycle to take into account possible 
screening effects.
We also performed such standard temperature-dependent DFT calculations and
found that the differences between the predictions of both approaches
were very small (the main effect was that the electronic entropies in both 
approaches had to be scaled).
For this reason and also because it is not clear that the self-consistent 
approach always leads to better results we used the computationally less
time-consuming non-self-consistent treatment of Eq.\ (\ref{eq_nscf}).
Our results for different laser-induced initial electronic temperatures are 
summarized in table \ref{table_coef}.
\begin{table*}
  \caption{\label{table_coef}
  Coefficients for the potential energy surfaces of Eq.\ (\ref{eq_surface})
  as a function of the electronic entropy $S_e$.
  In Eq.\ (\ref{eq_surface}) $E_\mathrm{tot}$ is given in mRy / atom,  
  $x$ and $y$ are given in units of the lattice parameter $a$, 
  and $z$ is given in units of $c$.
  The electronic temperature $T_e$ and the number of electron-hole pairs
  $N_{e-h}$ (expressed in $\%$ of the valence electrons), which are not 
  constants of motion, are given for $z = 0.234$ $c$.}
  \begin{ruledtabular}
  \begin{tabular}{ccccccccccc}
  $S_e$ ($k_B$ / atom) & $T_e$ (mRy) & $N_{e-h}$ ($\%$)
  & $a_0$   & $a_1$ ($10^4$) & $a_2$ ($10^8$) & $a_3$ ($10^{11}$)
  & $a_4$ ($10^{14}$) & $b_0$ ($10^3$) & $b_1$ ($10^7$) & $b_2$ ($10^{10}$) \\
  \hline
  $0.164$ & $12.7$ &
  $0.50$&$-12.087$ & $-3.0663$ & $1.2609$ & $-2.1728$ & $1.7787$ & $-3.35$
                                                      & $4.202$ & $-7.48$ \\
  $0.260$ & $16.2$ &
  $0.81$&$-11.428$ & $-2.4474$ & $1.0267$ & $-1.6487$ & $1.3035$ & $-2.52$ 
                                                      & $3.360$ & $-5.30$ \\
  $0.300$ & $17.6$ &
  $0.95$&$-11.039$ & $-2.2087$ & $0.9417$ & $-1.4619$ & $1.1342$ & $-2.21$ 
                                                      & $3.070$ & $-4.58$ \\
  $0.428$ & $21.9$ &
  $1.41$& $-9.392$ & $-1.5468$ & $0.7313$ & $-1.0278$ & $0.7519$ & $-1.36$
                                                      & $2.392$ & $-2.99$ \\
  $0.431$ & $22$   &
  $1.42$& $-9.346$ & $-1.5328$ & $0.7272$ & $-1.0200$ & $0.7452$ & $-1.34$
                                                      & $2.380$ & $-2.96$ \\
  \end{tabular}
  \end{ruledtabular}
\end{table*}

In the electronic ground state the total energy $E_\mathrm{tot}$ was minimized 
for $z = 0.2346$ $c$.
At this parameter the $A_{1g}$ and $E_g$ phonon frequencies were 
$2.89$ and $1.94$ THz, respectively, in reasonable agreement with experiment 
[$z = 0.2341$ $c$, $\nu( A_{1g}) = 2.92$, and $\nu( E_{g}) = 2.16$ THz] 
\cite{Hase02} and with earlier calculations [$z = 0.2336$ $c$, 
$\nu( A_{1g}) = 2.93$ THz] \cite{Murray05}.
The differences with the experimental results are probably due to the local
density approximation \cite{Perdew92}, which we used for the exchange and
correlation energy.
The differences with the earlier calculations \cite{Murray05} are probably due
to the Bi pseudopotential used in those calculations.
The present all-electron calculations do not rely on a pseudopotential and 
should be more accurate.

\begin{figure}
  \includegraphics[angle=270,width=8.0cm]{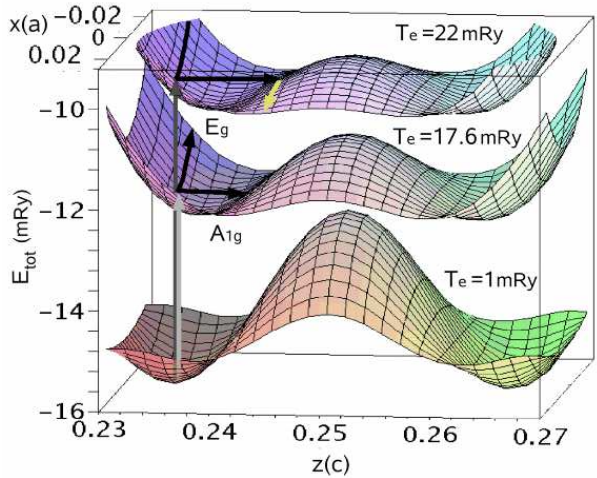}
  \caption{\label{fig_etot}(Color online)
  Potential energy surfaces for the ground state ($T_e = 1$ mRy) and for 
  excited electronic states ($T_e = 17.6$ and $22$ mRy at $z = 0.234$ $c$).
  Vertical arrows show transitions of the atoms from the ground state to an
  excited energy surface.
  The $A_{1g}$ and $E_g$ phonon modes are indicated.
  On the surface labeled $T_e = 22$ mRy the curvature in the $x$ direction 
  becomes negative when the atoms are at their maximal displacement in the 
  $z$ direction.
  This negative curvature is indicated by a light (yellow online) arrow.}
\end{figure}

In the excited electronic states (table \ref{table_coef}) the potential energy
surfaces are flatter than the ground state potential energy surface 
(Fig.\ \ref{fig_etot}). 
As a consequence the phonon frequencies are lower.
For large amplitude coherent phonons anharmonicity may further
lower the phonon frequencies \cite{Hase02}.
As mentioned above, in \cite{Sokolowski03} the maximal laser-induced 
displacement $\Delta z$ has been accurately determined:
Under the influence of a laser pulse that leads to a softening of
the $A_{1g}$ frequency from $2.92$ to $2.12$ THz, $\Delta z = 0.0085 \pm 
0.0021$ $c$.
For the electronic entropy $S_e = 0.428$ $k_B / \mathrm{atom}$ we reproduced
this softening of the $A_{1g}$ phonon frequency.
According to our calculations $16\%$ of the decrease of the $A_{1g}$ phonon 
frequency was due to the anharmonicity of the potential (The harmonic 
frequency was $2.24$ THz).
Our value for the amplitude of the $A_{1g}$ motion was $\Delta z = 0.0075$ 
$c$, in agreement with the experimental value \cite{Sokolowski03}. 

After the initial softening of the $A_{1g}$ phonons, the frequency returns to 
its original value within roughly $10$ ps \cite{Hase02}.
We discuss the origin of this frequency hardening by comparing our results and
the experimental results of \cite{Hase02}, where the $A_{1g}$ phonon frequency 
has been resolved in time.
For the electronic entropy $S_e = 0.300$ $k_B / \mathrm{atom}$ ($T_e = 17.6$
mRy at $z=0.234$ $c$) our calculated $A_{1g}$ phonon frequency was $2.45$ THz, 
the initial $A_{1g}$ frequency obtained in \cite{Hase02}.
After $0.3$ ps the $A_{1g}$ frequency in \cite{Hase02} increased to $2.52$ THz.
Our results indicated that at most $0.03$ THz of this increase can be explained
by the anharmonicity of the potential energy surface
(The harmonic frequency was $2.48$ THz),
in agreement with experiments using two time-delayed pump laser pulses 
\cite{Murray05} that have shown that the frequency of the $A_{1g}$ phonon mode
averaged over five periods is the same within $1\%$ independent of the 
amplitude of the oscillations for frequencies as low as $2.65$ THz.
Note however that anharmonicity plays a substantial role in the phonon
softening in \cite{Sokolowski03} as discussed above.
Instead a decrease of the electronic entropy to 
$S_e = 0.260$ $k_B / \mathrm{atom}$ yielding an $A_{1g}$ frequency of 
$2.52$ THz is probably responsible for the frequency increase observed in 
\cite{Hase02}.
The accompanying decrease in the number of electron-hole pairs by $15\%$ 
(calculated at $z = 0.234$ $c$) corresponds to a decay time of $1.8$ ps, which 
agrees well with the experimentally determined electronic background decay 
time of $1.78 \pm 0.08$ ps, which is independent of the laser fluence 
\cite{DeCamp01}.
The loss of entropy of the electrons near the surface is most likely due to 
diffusion of hot electrons away from the surface region (the penetration depth
of the laser $d \approx 17$ nm) and may also be partly due to the exchange of 
heat of the electrons with the ions via electron-phonon coupling, two 
processes that were not explicitly taken into account in our calculations.

At this point it is necessary to discuss  the limitations of the 
two-chemical potential model proposed in \cite{Tangney99}.   
According to \cite{Hase04} the electron-hole recombination time in Bi at 
$300$ K is about $1$ ps.
In a highly excited electronic state this time is expected to be considerably 
shorter.
So within less than four phonon periods the electrons and the holes obtain a 
common chemical potential.
To check the validity of the two-chemical-potential model we applied it and 
found that the chemical potential of the holes was greater than the chemical 
potential of the electrons, which means that upon reaching a common chemical 
potential the number of electron-hole pairs would increase, which would then 
give rise to a substantial further lowering of the phonon frequencies (we 
estimated that the ions would even go over the barrier at $z = 0.25$ $c$), in 
contradiction with experiment \cite{Hase02}.
Therefore, we believe that the two-chemical potential model \cite{Murray05}
does not provide a realistic description of the electron dynamics in Bi.
We can however not exclude that this model might be appropriate for 
semiconductors or for Bi at lower laser fluences.

\begin{figure}
  \includegraphics[angle=270,width=8cm]{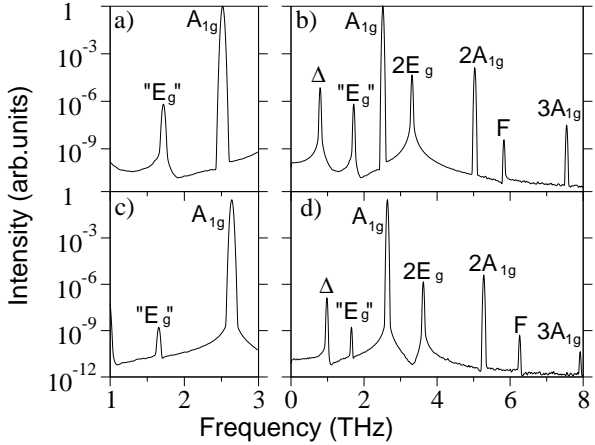}
  \caption{\label{fig_ft} 
  Fourier transforms (intensities) of the $z$ coordinate of the atoms for two 
  different simulations: a), b) $T_e = 16.2$ and c), d) $T_e = 12.7$ mRy 
  at $z=0.234$ $c$.
  All curves have been convoluted with a Gaussian with a full width at half
  maximum of $0.03$ THz.
  The left panels a) and c) show that the height of the peak labeled ``$E_g$'' 
  depends very sensitively on the laser fluence, which was twice lower in
  c) than in a).
  Experimentally this has been shown in \cite{Hase02}.}
\end{figure}

We now consider the coupling between the $A_{1g}$ and the $E_g$ phonons.
Figures \ref{fig_ft}(a) and \ref{fig_ft}(b) show the intensity of the 
Fourier transform of the $z$ coordinate of the atoms.
Higher harmonics of the main $A_{1g}$ peak at $2.52$ THz, which have
also been observed experimentally \cite{Misochko03}, and which are a 
consequence of the anharmonicity of the potential, are indicated.
For the initial velocity of the atoms in the $x$ direction (coherent $E_g$
phonons) we chose a value that gave a peak to peak amplitude 
$\Delta x = 0.4 \Delta z$, in agreement with \cite{Renucci73}.
In Fig.\ \ref{fig_ft}(b) it is clear from the peaks that are not higher 
harmonics of the main frequency that there is a considerable coupling between 
the $A_{1g}$ and $E_g$ phonons.
In Eq.\ (\ref{eq_surface}) the $A_{1g}$ phonon mode couples to $x^2 + y^2$,
a signal with double the $E_g$ frequency, $2 \nu(E_g)$.
Accordingly an analysis of the $x$ coordinate of the atoms showed that the 
frequency of the highest peak induced by the phonon-phonon coupling,
labeled $2 E_g$ in Fig.\ \ref{fig_ft}(b), ($3.32$ THz) equaled $2 \nu(E_g)$.
Experimentally this peak has been observed at $3.44$ THz \cite{Hase_private}.
So in our calculations $\nu(E_g)$ was slightly lower than in the
experiment \cite{Hase_private}, which is consistent with our ground state 
calculation.
The frequency of the peak labeled ``$E_g$'' in Fig.\ \ref{fig_ft}(b) 
($1.72$ THz) is given by $2 \nu(A_{1g}) - 2 \nu(E_g)$.
Experimentally this peak has been observed at $1.61$ THz \cite{Hase02}.
Our value was slightly higher than in the experiment \cite{Hase02} because
the calculated $\nu(E_g)$ was slightly lower.
Note that in \cite{Hase02} this peak has been identified with the $E_g$ mode.
However, since there is no coupling term linear in $x$ ($y$), and since the
$E_g$ mode produces modulations of the $A_{1g}$ oscillations, it is clear
that no peak of the power spectrum can have the frequency $\nu(E_g)$, whereas
the difference $2 \nu(A_{1g}) - 2 \nu(E_g)$ should be present.
The fact that the ``$E_g$'' peak has an intensity of only
$I($''$E_g$''$) = 7 \, 10^{-7} I(A_{1g})$ and that it has nevertheless been 
observed
experimentally \cite{Hase02} indicates that the laser-induced amplitude
of the $E_g$ mode is probably larger than $\Delta x = 0.4 \Delta z$, which we
used in our simulation. 
According to \cite{Renucci73} the relative Raman cross sections of the 
$A_{1g}$ and $E_g$ modes are indeed temperature dependent.
The sensitivity of the results to the initial conditions is further underlined
by a calculation with $\Delta x = 1.2 \Delta z$, for which we found 
$I($''$E_g$''$) = 3 \, 10^{-4} I(A_{1g})$.
Finally, we note that
the peak labeled $\Delta$ in Fig.\ \ref{fig_ft}(b) [$0.80$ THz 
$= 2 \nu(E_g) - \nu(A_{1g})$] is experimentally hard to observe because of
broadened spectral structures near zero frequency \cite{Hase_private}.
Similarly the peak labeled $F$ in Fig.\ \ref{fig_ft}(b) 
[$2 \nu(E_g) + \nu(A_{1g})$] probably has too low an intensity to be observed.

Figures \ref{fig_ft}(c) and \ref{fig_ft}(d) were calculated for the case where
the energy absorbed from the laser was twice as small as in 
Figs.\ \ref{fig_ft}(a) and \ref{fig_ft}(b) 
($T_e = 12.7$ mRy at $z = 0.234$ $c$).
It shows that the intensity of the peak ``$E_g$'' became orders of magnitude 
lower upon reducing the laser fluence by a factor of two.
Therefore a strong laser pulse is needed to observe the $A_{1g}$-$E_g$ phonon 
coupling, which has also been found experimentally in \cite{Hase02}.

Another interesting result that we found is that the coupling between the 
$A_{1g}$ and $E_g$ modes became very strong for 
$S_e \agt 0.431 k_B / \mathrm{atom}$ ($T_e = 22$ mRy at $z = 0.234$ $c$):
In this case the curvature of the potential energy surface in the
$x$ direction became negative every time $z$ reached its maximum value
(Fig.\ \ref{fig_etot}).
We found that the amplitude of the $E_g$ mode increased 
exponentially even when the initial velocity in the $x$ direction was
very small.

In conclusion, we presented a dynamical simulation of coherent
laser-induced $A_{1g}$ and $E_g$ phonons in Bi on a potential energy surface 
that was obtained by all-electron, \textit{first principles} calculations.
Using these simulations we could explain the effects of anharmonicity
and $A_{1g}$-$E_g$ phonon coupling that take place after the interaction with 
ultrashort, intense laser pulses.
We found good agreement with all experiments performed so far 
\cite{Sokolowski03,Hase02,Hase_private,Murray05,Hase96,Misochko03}.

This work has been supported by the Deutsche Forschungsgemeinschaft (DFG) 
through the priority program SPP 1134 and by the European Community Research 
Training Network FLASH (MRTN-CT-2003- 503641).


\end{document}